\documentclass{article}

\usepackage{arxiv}
\usepackage{amssymb}
\usepackage{amsmath}
\usepackage{amscd}
\usepackage{graphicx}
\usepackage{graphics}
\usepackage{xcolor,graphicx}
\usepackage{csquotes}
\usepackage{mathtools}
\usepackage{caption}
\usepackage{float}
\usepackage{subcaption}
\usepackage{cite}
\usepackage{multirow}
\usepackage{array}
\usepackage{varwidth} % must be loaded after array
\usepackage{enumitem }
\usepackage{algorithm}
\usepackage{algorithmic}
\usepackage{booktabs}
\usepackage[defaultcolor=magenta]{changes}
\definechangesauthor[name={Per cusse}, color=orange]{per}

\usepackage{nicefrac}
\usepackage{optidef}
\usepackage{hyperref}
\hypersetup{colorlinks,linkcolor={blue},citecolor={blue},urlcolor={red}}
\usepackage{pbox}
\usepackage{lettrine}
\renewcommand{\bold}[1]{\mathbf{#1}}

\newcommand{\diag}{\text{diag}}

\renewcommand{\S}[1]{\bold{S}({#1})} 
\DeclareMathOperator*{\minimize}{minimize}

\renewcommand{\P}{\bold{P}}

\renewcommand{\u}{\bold{u}}
\renewcommand{\d}{\bold{d}}
\renewcommand{\b}{\bold{b}}
\newcommand{\m}{\bold{m}}
\newcommand{\A}{\bold{A}}

\renewcommand{\S}{\bold{S}}

\renewcommand{\H}{\bold{H}}

\newcommand{\Eps}{\boldsymbol{\varepsilon}}
\newcommand{\bNu}{\boldsymbol{\nu}}
\newcommand{\lamb}{\boldsymbol{\lambda}}

\usepackage{relsize}

\begin{document}
%
% paper title
% can use linebreaks \\ within to get better formatting as desired
\title{Fast and Automatic Full Waveform Inversion by Dual Augmented Lagrangian}

%
%
% author names and IEEE memberships
% note positions of commas and nonbreaking spaces ( ~ ) LaTeX will not break
% a structure at a ~ so this keeps an author's name from being broken across
% two lines.
% use \thanks{} to gain access to the first footnote area
% a separate \thanks must be used for each paragraph as LaTeX2e's \thanks
% was not built to handle multiple paragraphs
%

\author{
 Kamal Aghazade \\
Institute of Geophysics, Polish Academy of Sciences, Warsaw, Poland. \\
 \texttt{aghazade.kamal@igf.edu.pl} \\
  %% examples of more authors
   \And
 Ali Gholami \\
  Institute of Geophysics, Polish Academy of Sciences, Warsaw, Poland, \\
  \texttt{agholami@igf.edu.pl}
}

\maketitle

\begin{abstract}
Full Waveform Inversion (FWI) stands as a nonlinear, high-resolution technology for subsurface imaging via surface-recorded data. This paper introduces an augmented Lagrangian dual formulation for FWI, rooted in the viewpoint that Lagrange multipliers serve as fundamental unknowns for the accurate linearization of the FWI problem. Once these multipliers are estimated, the determination of model parameters becomes simple. Therefore, unlike traditional primal algorithms, the proposed dual method circumvents direct engagement with model parameters or wavefields, instead tackling the estimation of Lagrange multipliers through a gradient ascent iteration. This approach yields two significant advantages: i) the background model remains fixed, requiring only one LU matrix factorization  for each frequency inversion. ii) Convergence of the algorithm can be improved by leveraging techniques like quasi-Newton l-BFGS methods and Anderson acceleration. Numerical examples from elastic and acoustic FWI utilizing different benchmark models are provided, showing that the dual algorithm converges quickly and requires fewer computations than the standard primal algorithm.
\end{abstract}

%% Note that keywords are not normally used for peerreview papers.
%\begin{IEEEkeywords}
%Full Waveform Inversion, Extended-source FWI, Lagrange multipliers, Augmented Lagrangian, dual methods. 
%\end{IEEEkeywords}

\graphicspath{{"./figures/"}}
\section{Introduction}

Full waveform inversion (FWI) is a widely used seismic imaging technique with a focus on reconstruction of the physical properties of the subsurface (e.g., velocity, density, attenuation, conductivity, permittivity, and anisotropic parameters) \cite{Virieux_2009_overview,aghamiry2019compound}. It has a broad application in various field of geoscience such as global seismology (see e.g., \cite{Tromp_2019_SWI}, and references therein), ground penetration radar (GPR) \cite{feng2021wavefield,giannakis2018realistic}, CO$_2$ characterization \cite{Dupuy_2017_CO2}, glaciology \cite{Pearce_2023_CIS},
near surface studies by inversion of surface waves \cite{Wittkamp_2018_JRL}, geothermal studies \cite{Qu_2024_DAS},  %\cite{Kasahara_2019_DAS}, 
and volcanology \cite{Klaasen_2024_DAS}.

\begin{figure*}[!t]
\centering
 \includegraphics[width=.95\columnwidth]{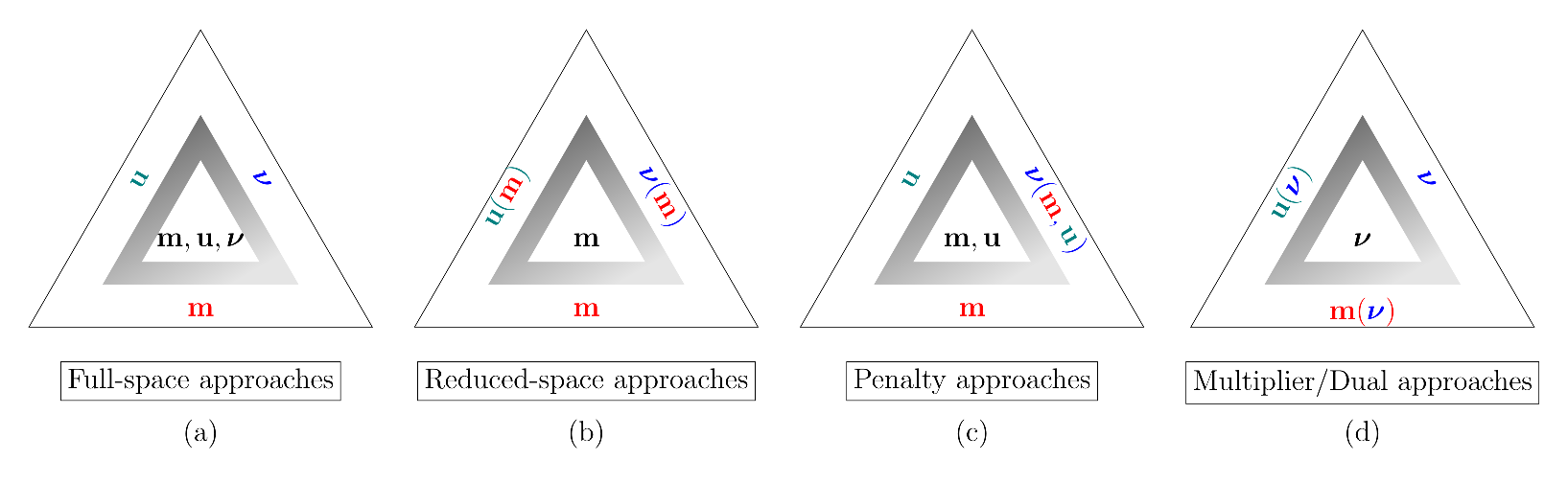}
\caption{Different FWI formulations: (a) Full-space approach, which treats $\m$, $\u$, and $\bNu$ as optimization parameters; (b) Conventional FWI formulation, which eliminates the wavefield and Lagrange multipliers from the optimization variables; (c) Penalty formulation, which eliminates only the Lagrange multipliers from the optimization variables; (d) The approach proposed in this study, which eliminates the model parameters and wavefield from the optimization variables.}
\label{FWI_methods}
 \end{figure*}

Even though FWI can be performed in either the time or frequency domain, this paper focuses on frequency domain FWI \cite{Pratt_1998_GNF}.
\subsection{Forward Problem} \label{forward_sec}
\subsubsection{Acoustic media}
For a given medium characterized by the model parameters vector $\m \in \mathbb{R}^{n \times 1}$ (squared slowness), the data vector $\d \in \mathbb{C}^{n_r \times 1}$ is calculated by first solving the Helmholtz equation 
\begin{equation}\label{WE_ac}
  (\omega^2 \text{diag}(\m)+\Delta)\u  = \b,
\end{equation}
for the wavefield $\u \in \mathbb{C}^{n \times 1}$. The monochromatic wavefield $\u$ is then sampled at the receiver locations using the sampling operator $\P \in \mathbb{R}^{n_r \times n}$, i.e., $\d = \P \u$.
In these equations, $\omega$ denotes angular frequency, $\Delta$ is Laplace operator, $\b \in \mathbb{C}^{n \times 1}$ is the source term, $n$ is the number of model parameters, and $n_r$ denotes the number of receivers.

Equation \eqref{WE_ac} can be discretized following the methodology described in  \cite{Chen_2013_optimal_FD}.
The discretized equation can be written in matrix-vector form as $\A(\m)\u  = \b$. Here, $\A(\m)=(\omega^2 \text{diag}(\m)+\Delta) \in \mathbb{C}^{n \times n}$ represents the Helmholtz operator, which is constructed with sufficient accuracy and suitable boundary conditions. Following the bi-linearity of the wave equation, it is also possible to write \eqref{WE_ac} as $\bold{L}(\u)\m=\bold{y}(\u)$, where $\bold{L}(\u) = \omega^2 \text{diag}(\u)$, and $\bold{y}(\u) = \b-\Delta \u$.

%
%
% \subsubsection{Elastic media}
% The elastic wave equation can be defined as:

\subsubsection{Elastic media}
The elastic wave equation can be defined as:
\begin{equation}
\begin{aligned}
\label{El_forward}
\small
&\boldsymbol{\rho} \omega^{2}\u_x\!\!+\!(\boldsymbol{\lambda}\!+\!2\boldsymbol{\mu})\partial_{xx} \u_x\!\!+\!\boldsymbol{\mu} \partial_{zz} \u_x \! +\! (\boldsymbol{\lambda}+\boldsymbol{\mu})\partial_{xz}\u_z \!= \!\b_x \\
&(\boldsymbol{\lambda}+\boldsymbol{\mu})\partial_{xz} \u_x + \boldsymbol{\rho} \omega^{2} \u_z\!\!+\!(\boldsymbol{\lambda}\!+\!2\boldsymbol{\mu})\partial_{zz} \u_z\!\!+\!\boldsymbol{\mu} \partial_{xx}\u_z \!=\! \b_z
\end{aligned}
\end{equation}
%
%
% \begin{align}\label{El_forward}
% &
% \begin{pmatrix}
%     \boldsymbol{\rho} \omega^{2}\!\!+\!(\boldsymbol{\lambda}\!+\!2\boldsymbol{\mu})\partial_{xx}\!\!+\!\boldsymbol{\mu} \partial_{zz} & (\boldsymbol{\lambda}+\boldsymbol{\mu})\partial_{xz} \\
%     (\boldsymbol{\lambda}+\boldsymbol{\mu})\partial_{xz} & \boldsymbol{\rho} \omega^{2}\!\!+\!(\boldsymbol{\lambda}\!+\!2\boldsymbol{\mu})\partial_{zz}\!\!+\!\boldsymbol{\mu} \partial_{xx}
% \end{pmatrix} \nonumber
% \\
% &~~~~~~~~~~~
% \begin{pmatrix}
%      \bold{u}_x\\
%      \bold{u}_z
% \end{pmatrix}=
% \begin{pmatrix}
%      \bold{b}_x\\
%      \bold{b}_z
% \end{pmatrix}
% \end{align}

where $\boldsymbol{\rho} \in \mathbb{R}^{n \times 1}$ is mass density, $\boldsymbol{\lambda} \in \mathbb{R}^{n \times 1}$ and $\boldsymbol{\mu} \in \mathbb{R}^{n \times 1}$ denote Lamé parameters,
 $\bold{u}_x \in \mathbb{C}^{n \times 1}$ and $\bold{u}_z \in \mathbb{C}^{n \times 1}$ are horizontal and vertical particle displacements, and $\bold{b}_x \in \mathbb{C}^{n \times 1}$, $\bold{b}_z \in \mathbb{C}^{n \times 1}$ are the source terms. 
Using the definition of P- and S- wave velocities: 
\begin{equation*}
    \text{V}_{\text{P}} = \sqrt{\frac{\boldsymbol{\lambda}+2\boldsymbol{\mu}}{\boldsymbol{\rho}}},
    \quad 
    \text{V}_{\text{S}} = \sqrt{\frac{\boldsymbol{\mu}}{\boldsymbol{\rho}}},
\end{equation*}
and assuming a constant density medium, then \eqref{El_forward} can be discretized and written in matrix-vector form as 
$\A(\m)\u = \b$
where 
\begin{equation}
 \m = 
\begin{pmatrix} 
 \m_\text{P}:=\text{V}_{\text{P}}^2\\
 \m_\text{S}:=\text{V}_{\text{S}}^2
\end{pmatrix},
\quad
\bold{u} = 
\begin{pmatrix} 
\u_x\\
\u_z
\end{pmatrix},
\quad
\b = 
\begin{pmatrix} 
\bold{b}_x\\
\bold{b}_z
\end{pmatrix}.
\end{equation}
In this case, $\A(\m) \in \mathbb{C}^{2n \times 2n}$ is the elastic forward operator, $\u \in \mathbb{C}^{2n \times 1}$ and $\b \in \mathbb{C}^{2n \times 1}$ includes the vertical and horizontal components of the wavefield and source, respectively. Also, $\d = \P \u \in \mathbb{C}^{2n_r \times 1}$ is the recorded data and $\P \in \mathbb{R}^{2n_r \times 2n}$ is the sampling operator.
The employed squared velocity parametrerization also allows writing \eqref{El_forward} as  $\bold{L}(\u)\m=\bold{y}(\u)$, where  $\bold{L}(\u) = \frac{\partial \A(\m)}{\partial \m}\u$ is defined as
 \begin{equation*}
      \bold{L}(\u) \! = \! \! \ \begin{bmatrix}
\diag (\partial_{xx}\bold{u}_{x} \! + \!\partial_{xz}\bold{u}_z) \! \! &  \!\diag(\partial_{zz}\bold{u}_{x} \!-\!\partial_{xz}\bold{u}_z)  \\
\diag(\partial_{zz}\bold{u}_z \! + \! \partial_{xz}\bold{u}_{x}) & \diag(\partial_{xx}\bold{u}_z \! - \! \partial_{xz}\bold{u}_{x})
\end{bmatrix}, 
 \end{equation*}
 and $\bold{y}(\u)=\b-\omega^2 \u$.
\subsection{Inverse Problem}
Mathematically, for the given source term $\b$ and the observed data $\d$, the inverse problem is to estimate the model parameters $\m$. This can be defined as a non-linearly constrained optimization problem, thorough which a data misfit term is minimized while simultaneously the wave equation is satisfied \cite{Haber_2000_OTS}:
\begin{equation}\label{FWI_const}
\minimize_{\m,\u}~ \frac{1}{2}\|\P \u-\d\|_{2}^{2}\quad \text{subject to} \quad \A(\m)\u=\b.
\end{equation}
The well-established method of Lagrange multipliers can be used to solve the inverse problem \cite{Nocedal_2006_NO}. This approach involves optimization of the Lagrangian function, defined as
\begin{equation} \label{Lagrangian}
     \mathcal{L}(\m,\u,\bNu) = \frac{1}{2}\|\P\u-\d\|_{2}^{2}+ \bNu^{T}(\A(\m)\u-\b),
\end{equation}
where $\bNu$ is the dual variable (or Lagrange multipliers associated to the wave-equation constraints). $\bold{x}^T$ denote the complex conjugate transpose of vector $\bold{x}$. The optimum point of the Lagrangian is a saddle-point which attains a minimum over the primal variables 
 $\m$ and $\u$ and a maximum over the dual variable $\bNu$. 
Different approaches have been proposed to solve the problem and find the optimum point (see Fig. \ref{FWI_methods}). 

%\begin{equation}\label{proximal_LF}
%\hat{\mathcal{L}}(\m,\u,\bNu) = \mathcal{L}(\m,\u,\bNu) -\frac{1}{\mu}\mathcal{R}(\bNu),
%\end{equation}
%where

%
%
%is the classical Lagrangian function associated with the constrained problem in \eqref{FWI_const}. In mathematical terminology, $\u$ and $\m$ are referred to as primal variables, while $\bNu \in \mathbb{C}^{N \times 1}$ represents the vector of Lagrange multipliers (or dual variable). 
 %
 %
% $\mathcal{R}(\bNu)$ is a proximal term (a regularization term) to stabilize the optimization process, and $\mu >0$ is the regularization parameter. In all-at-once strategy, ($\u$, $\m$, $\bNu$) are updated simultaneously (Fig.~\ref{FWI_methods}) by solving a memory intensive and ill-conditioned system of equations \cite{Haber_2000_OTS,Nocedal_2006_NO}, which is not suitable for FWI applications.
 %(also called KKT system, \cite{Nocedal_2006_NO}).

\subsubsection{Full-space approaches}
In the so-called full-space approach, the optimization is performed simultaneously over the variables $\m$, $\u$, and $\bNu$ (Fig. \ref{FWI_methods}a). Two strategies can be adopted for this approach. The first is the so-called all-at-once method, where the optimization is conducted jointly over all three variables \cite{Akcelik_2002_MNK}. This is typically achieved using a Newton-Krylov method, which solves the associated optimality conditions. Alternatively, a sequential optimization strategy can be employed, where the Lagrangian or augmented Lagrangian (AL) is optimized with respect to one of the variables $\m$, $\u$, and $\bNu$ at a time, while the other two variables are held constant \cite{Aghamiry_2019_IRWRI,Gholami_2022_EFW,Operto_2023_EFI}.

\subsubsection{Reduced-space approaches}
To address the memory complexity associated with handling all variables, a second group of methods eliminates the wavefield and multipliers from the set of optimization variables through variable projection \cite{golub2003separable}, thereby satisfying the wave equation constraint exactly. This approach expresses the wavefield and multipliers as implicit functions of $\m$, i.e.,  $\u(\m), \bNu(\m)$ (Fig. \ref{FWI_methods}b), leading to a more complicated objective function that depends only on the model parameters. As a result, the dimensionality of the search space is reduced. However, the reduced objective function is highly sensitive to the initial model and is challenging to optimize due to ill-conditioning and difficulties in tuning the step length \cite{Virieux_2009_overview}. Considerable effort has been made to improve the reduced FWI (e.g., \cite{VanLeeuwen_2010_CMC,Luo_2011_DBO,Warner_2016_AWI,Engquist_2016_WAS,Ha_2021_DBO,HE_2023_OMF}).
%. These include for example, cross-correlation \cite{VanLeeuwen_2010_CMC}, deconvolution \cite{Luo_2011_DBO,Warner_2016_AWI}, and optimal transport distance \cite{Engquist_2016_WAS,Metivier_2016_OTA}. All of these metrics belong to the category of reduced approach, where their implementation still faces other challenges.

\subsubsection{Penalty approaches}
One approach to address the ill-conditioning issue of the reduced FWI is to include a damping term in the Lagrangian, which penalizes the Lagrange multipliers. This method allows us to compute the Lagrange multipliers as a closed-form function of the model parameters and the wavefield, denoted as $\bNu(\m, \u)$. Consequently, this enables their elimination from the optimization variables, resulting in the quadratic penalty formulation of the FWI \cite{VanLeeuwen_2013_MLM}. The penalty approach in FWI requires optimization of the penalty function over $\m$ and $\u$ (Fig.~\ref{FWI_methods}c). This optimization can be performed through an alternating optimization method known as wavefield reconstruction inversion (WRI) \cite{VanLeeuwen_2013_MLM}. Additionally, it is possible to further eliminate the associated wavefields to construct a more complex reduced penalty objective function, which is then optimized solely over $\m$ \cite{van2015penalty}.
For a comprehensive review of penalty-based FWI, see \cite{Operto_2023_EFI}. However, the implementation of WRI suffers from inaccurate estimation of the Lagrange multipliers, which can significantly reduce the performance of the method, particularly when the initial model is erroneous \cite{Gholami_2024_GJI}.

%formulate the problem in the frame of proximal Lagrangian, as delineated in \cite{Gholami_2024_GJI}, wherein the standard Lagrangian function incorporates a quadratic proximal term to regularize the subproblem pertaining to the Lagrange multipliers throughout each iteration, thereby stabilizing the optimization process \cite{Rockafellar_1976_MOA}.

%wavefield a In this formulation, FWI is framed as an iterative, two-stage procedure. The initial stage entails solving a convex-concave min-max optimization problem to determine the wavefields and multipliers. The subsequent stage involves updating the model parameters in an analogous manner to standard FWI via correlation of the computed wavefields and multipliers. Manipulation of a 2 by 2 block linear saddle point system is required for resolving the initial stage. 

A more advanced FWI algorithm can be achieved through the proximal regularization of the Lagrangian \cite{Rockafellar_1976_MOA}, which regularizes the Lagrange multipliers by keeping them close to their previous estimates rather than damping them to zero, as done in penalty-based methods \cite{Gholami_2024_GJI}. This approach is equivalent to the standard AL-based FWI \cite{Aghamiry_2019_IRWRI}. In this method, optimization is performed over all three classes of parameters: $\m$, $\u$, and $\bNu$ (Fig. \ref{FWI_methods}a).
The proximal regularization employed by the AL method significantly enhances the performance of the FWI algorithm, particularly shown on synthetic data \cite{Aghamiry_2019_IRWRI,Aghazade_2021_SWI}.
 
\subsection{Contributions}
In this paper, we develop, for the first time, a dual approach for FWI, where the model parameters and wavefield are expressed as implicit functions of the Lagrange multipliers, $\m(\bNu)$ and $\u(\bNu)$ (Fig. \ref{FWI_methods}d). This formulation allows us to eliminate $\m$ and $\u$ from the optimization variables, leading to a dual AL function dependent solely on $\bNu$.
The proposed dual approach reduces the FWI problem to the estimation of accurate Lagrange multipliers by solving the dual objective function. Once the multipliers are estimated, the FWI problem can be resolved efficiently in a single iteration.

The dual function is maximized using a gradient ascent algorithm, which can be further accelerated by applying the limited memory BFGS (l-BFGS) \cite{Liu_1989_LMB} or Anderson Acceleration (AA) techniques \cite{Walker_2011_AAF, Aghazade_2022_AAA}. A main advantage of this dual formulation is that for each frequency, the associated multiplier is estimated based on the initial model for that frequency. Consequently, the wave equation operator is dependent on the initial model and remains fixed during the iteration. Therefore, only a single lower-upper triangular (LU) decomposition is necessary to invert each data frequency, akin to the standard contrast source inversion method \cite{Abubakar_2009_CSI}. However, while the standard contrast source inversion is based on the penalty formulation of FWI, the proposed algorithm utilizes the AL formulation. Thus, it combines the computational efficiency of contrast source inversion with the accuracy of the AL algorithm.

The proposed algorithm is tested using both acoustic and elastic data. The results demonstrate that accurate results can be obtained efficiently and fully automatically.

Section \ref{dual_method} presents the theory behind the dual AL method. In Section \ref{Alg_des}, the algorithm for implementing the dual AL method is detailed, including aspects such as acceleration techniques, parameter selection, and computational complexity. Section \ref{Interp} provides a comparison of the dual algorithm with traditional methods like FWI and WRI. Finally, Section \ref{NumExam} shows numerical experiments using both acoustic and elastic data.
\section{FWI by Dual AL method} \label{dual_method}
%\subsection{FWI from the perspective of Lagrange multipliers}
 
 The AL for the problem \eqref{FWI_const} is defined as
\begin{equation}
 \mathcal{L}_{\mu}(\m,\u,\bNu) = \mathcal{L}(\m,\u,\bNu)+\frac{\mu}{2}\|\A(\m)\u-\b\|_{2}^{2},
\end{equation}
where $\mathcal{L}$ is the standard Lagrangian defined in \eqref{Lagrangian} and $\mu>0$ is the penalty parameter. 
Optimization of the AL function requires minimization of it with respect to the primal variables $\m$ and $\u$ and maximization over the Lagrange multiplier $\bNu$. 

The dual problem is defined as
\begin{equation}
 \max_{\bNu} \mathcal{D}_{\mu}(\bNu),
\end{equation}
where
\begin{equation}
\mathcal{D}_{\mu}(\bNu) =\min_{\m,\u} \mathcal{L}_{\mu}(\m,\u,\bNu).
\end{equation}
It can be shown that the gradient of the dual function has the following simple form (see \cite{tapia1977diagonalized}):
\begin{equation}
 \frac{\partial}{\partial \bNu} \mathcal{D}_{\mu}(\bNu) = \A(\m(\bNu))\u(\bNu)-\b,
\end{equation}
where $\m(\bNu)$ and $\u(\bNu)$ are the model parameters and the wavefield that are obtained by minimizing the AL function for a given multiplier vector $\bNu$. 

\subsection{Computation of $\m(\bNu)$ and $\u(\bNu)$}
The $\m(\bNu)$ and $\u(\bNu)$ are defined as
\begin{equation}
\arg\min_{\m,\u} \mathcal{L}_{\mu}(\m,\u,\bNu).
\end{equation}
It may not be possible to find closed-form expressions for $\m(\bNu)$ and $\u(\bNu)$ that minimize $\mathcal{L}_{\mu}$ in terms of $\bNu$. However, good approximations can be obtained in the neighborhood of an initial model $\m_0$.
%\subsubsection{Acoustic case}
For a given initial model $\m_0$, the wavefield $\u(\bNu)$ minimizing $\mathcal{L}_{\mu}$ satisfies
\begin{equation}
    \left( \P^T\P + \mu \A_0^T\A_0\right)\u(\bNu) = \P^T\d + \mu \A_0^T\b - \A_0^T\bNu.
\end{equation}
where $\A_0\equiv \A(\m_0)$.
It can be shown that the wavefield also satisfies \cite{Gholami_2024_GJI}
\begin{equation} \label{u_mult}
    \A_0\u(\bNu) = \b + \lamb(\bNu) - \frac{1}{\mu}\bNu,
\end{equation}
where 
\begin{equation}\label{lambda_mult}
    \lamb(\bNu) = \bold{S}_0^T(\bold{S}_0\bold{S}_0^T+\mu \bold{I})^{-1} (\delta \bold{d}_0 + \frac{1}{\mu}\bold{S}_0\bNu),
\end{equation}
with $\bold{S}_0=\P\A_0^{-1}$ and $\delta \bold{d}_0=\d - \bold{S}_0\b$.

Regarding the minimizer $\m(\bNu)$, %keeping in mind that $\A(\m)=\omega^2 \text{diag}(\m)+\Delta$,
%\begin{equation}
% \mathcal{L}_{\mu}(\m,\u,\bNu) = \frac{1}{2}\|\P\u-\d\|_{2}^{2}+ \bNu^{T}(\omega^2 \text{diag}(\m)\u+\Delta\u-\b)+\frac{\mu}{2}\|\omega^2 %\text{diag}(\m)\u+\Delta\u-\b\|_{2}^{2},
%\end{equation}
partial derivative of $\mathcal{L}_{\mu}$ with respect to $\m$ is 
\begin{align} \label{Lagrange_deriv}
    \frac{\partial \mathcal{L}_{\mu}}{\partial \m} =\hat{\bold{L}}(\bNu)^T\bNu + \mu \hat{\bold{L}}(\bNu)^T [\hat{\bold{L}}(\bNu)\m(\bNu) - \hat{\bold{y}}(\bNu)].
\end{align}
where $\hat{\bold{L}}(\bNu) \equiv \bold{L}(\u(\bNu))$ and $\hat{\bold{y}}(\bNu) \equiv \bold{y}(\u(\bNu)) $ (see Section \ref{forward_sec} for the definition of $\bold{L}$ and $\bold{y}$).
Using \eqref{u_mult} and the equality
\begin{equation*}
    \hat{\bold{L}}(\bNu)\m_0-\hat{\bold{y}}(\bNu)=\A_0\u(\bNu) - \b,
\end{equation*}
and letting $ \frac{\partial \mathcal{L}_{\mu}}{\partial \m} =0$, we get
\begin{equation} \label{m_m0}
\m(\bNu) =\m_0 - [\hat{\bold{L}}(\bNu)^T\hat{\bold{L}}(\bNu)]^{-1}\hat{\bold{L}}(\bNu)^T \lamb(\bNu).
\end{equation}

$
%\delta{\m}^{k}=-\frac{\text{real}(\langle \Lam^{k}, \u^{k}\rangle) }{\omega^2 \langle  \u^{k},  \u^{k}\rangle}.
$
%\STATE Update the secondary Lagrange multiplier
$
%\eps^{+}= \eps + \A(\m+\delta{\m})\up-\b)  
$
%\ENDFOR
%\STATE Use $\delta{\m}^{k}$ obtained at the current frequency to update the model, ${\m}\leftarrow \bold{m}+\delta{\m}^{k}$, and use it for the inversion at the next higher frequency.
%\ENDFOR
%\end{algorithmic}
%\end{algorithm}

\section{Algorithm Description} \label{Alg_des}
The dual algorithm presented above is closely related to standard AL algorithms for FWI \cite{Aghamiry_2019_IRWRI, Gholami_2022_EFW, Operto_2023_EFI, Gholami_2024_GJI}, with one key difference: in the dual algorithm, the background model remains fixed, and the focus shifts to estimating the associated multiplier. In this case, the stiffness matrix depends solely on the background medium, which remains unchanged throughout the inversion process. Consequently, when using a direct solver, such as the LU decomposition method, the finite-difference (FD) operator is factorized and the Hessian matrix is computed only once. These results can then be reused for multiple sources and across successive iterations of the inversion \cite{Abubakar_2009_CSI}.

Using a scaled Lagrange multiplier $\Eps=\frac{1}{\mu}\bNu$, the forward/backward wavefields $\u(\Eps)/\lamb(\Eps)$ can be computed using either of the following approaches \cite{Gholami_2024_GJI}:
\begin{enumerate}
    \item Wavefield-oriented approach:
\begin{align}
    \u(\Eps)\! &=\! \left( \P^T\P\! + \!\mu \A_0^T\!\A_0\right)^{\!-1}\!\!(\P^T\!\d + \!\mu \A_0^T(\b - \Eps)), \label{u_wo}\\
    \lamb(\Eps) &= \Eps+\A_0\u(\Eps) - \b.
\end{align}
\item Multiplier-oriented approach:
\begin{align}
    \lamb(\Eps) &= \bold{S}_0^T(\bold{S}_0\bold{S}_0^T+\mu \bold{I})^{-1} (\delta \bold{d}_0 + \bold{S}_0\Eps), \label{moalm_lambda}\\
    \u(\Eps) &= \A_0^{-1}(\b + \lamb(\Eps) - \Eps). \label{moalm_u}
\end{align}
\end{enumerate}
Then the gradient of the dual objective is simply $ \bold{g}(\Eps)=\A(\m(\Eps))\u(\Eps)-\b$, which allows us to solve the problem by the standard gradient based algorithms \cite{Nocedal_2006_NO}.
The steepest ascent method is the simplest algorithm for maximizing the dual objective function, as it only requires computing the gradient of the objective function at each iteration.
Algorithms \ref{alg1} and \ref{alg2} summarize respectively the multiplier-oriented and wavefield-oriented forms of the dual method for multi-frequency FWI.
Due to the high computational cost of computing the forward and backward wavefields at each iteration of the optimization, improving the algorithm's convergence rate can be of great significance. In the following, we describe two acceleration methods that require only the gradient to be supplied.

\subsection{Quasi-Newton l-BFGS method}
The update direction provided by the Newton’s method is
\begin{equation}
\bold{x}_k= \H^{-1}_k \bold{g}_k,
\end{equation}
where $k$ is the iteration number, $\bold{g}_k$ is the gradient at iteration $k$, and $\H_k$ is the Hessian matrix of the dual function at iteration $k$.
Computing the inverse of the Hessian at each iteration can be challenging. A quasi-Newton method iteratively approximates the Hessian matrix, with the l-BFGS method being one of the most popular. l-BFGS updates the Hessian approximation using changes in both the solution and the gradient at each iteration \cite{Nocedal_2006_NO}:
\begin{equation}
\H_{k+1}=\H_k + \frac{\delta\bold{g}_k\delta\bold{g}_k^T}{\delta \bold{g}_k^T\delta\Eps_k} - 
\frac{\H_k \delta\Eps_k\delta\Eps_k^T\H_k^T}{\delta\Eps_k^T\H_k\delta\bold{g}_k},
\end{equation}
where $\delta\Eps_k=\Eps_{k+1}-\Eps_k$ and $\delta \bold{g}_k= \bold{g}_{k+1}-\bold{g}_k$.
Having computed the  update direction $\bold{x}_k$, the multiplier is updated as
\begin{equation}
    \Eps_{k+1}=\Eps_k + \alpha_k \bold{x}_k,
\end{equation}
where $\alpha_k$ is the step length and is chosen to satisfy the Wolfe conditions.
Despite relying solely on the gradient, l-BFGS provides a sufficiently accurate approximation of the objective function to achieve superlinear convergence.

\subsection{Anderson acceleration} 
The gradient ascent algorithm for updating the multiplier may be defined as a fixed-point iteration:
\begin{equation} \label{FP}
    \Eps_{k+1}=g(\Eps_k):=\Eps_k + \A(\m(\Eps_k))\bold{u}(\Eps_k)-\bold{b}.
\end{equation}
This enables the use of acceleration strategies to improve convergence rates, such as Anderson Acceleration (AA) \cite{Walker_2011_AAF}. In this paper, we employ AA. For a detailed implementation of AA in solving \eqref{FP}, the reader is referred to \cite{Aghazade_2022_AAA}. In \cite{Aghazade_2022_AAA}, the AA method was applied to improve the convergence of the wavefield-oriented AL approach in acoustic FWI, where the search space includes not only the Lagrange multipliers but also the model parameters and the wavefield.
%%%%%%%%%%%%%%%%%%%%%%%%%%%%%%%%%%%%%%%%%%%%%%%%%%%%%%%%%%%%%%%%%%%%%%%%%%%%%%%%%%%%%%%

\begin{algorithm}[H]
 \begin{algorithmic}[1]
 \caption{Multiplier-oriented dual AL algorithm for FWI.} \label{alg1}
 \REQUIRE Observed data $\d$, source $\b$, initial model $\m$, fixed data tolerance $\delta_{\omega}$
 \FOR{$\omega\in[\omega_{min},\omega_{max}]$}
 \STATE Compute LU factorization of the Helmholtz operator $\A_0=\A(\m)$ at the frequency $\omega$
 \STATE  Form the data-space Hessian matrix $\bold{Q}=\bold{S}_0\bold{S}_0^T=\P\A_0^{-1}\A_0^{-T}\P^T$ 
 \STATE  Set $\Eps_0=\bold{0}$
  \FOR{$k=1,2,...,maxit$}
   \STATE  $\delta\bold{d}_{\omega}=\d_{\omega}-\S_0(\b_{\omega}-\Eps_{k-1})$
   \STATE  Solve $\|(\frac{1}{\mu} \bold{Q} +  \bold{I})^{-1}\delta\bold{d}_{\omega}\|_2= \delta_{\omega}$ for $\mu$
 \STATE  $\lamb=\S_0^T(\bold{Q} +  \mu\bold{I})^{-1}\delta\d_{\omega}$
\STATE  $\u =\A_0^{-1}(\b_{\omega} +\lamb -\Eps_{k-1})$
\STATE  $\delta \m=(\bold{L}(\u)^T\bold{L}(\u))^{-1}\bold{L}(\u)^T \lamb$
\STATE $\Eps_{k}=\Eps_{k-1} + \A(\bold{m}+\delta \m)\bold{u}-\bold{b}_{\omega}$
\ENDFOR
\STATE Use $\delta\m$ obtained at the current frequency to update the model, $\m \leftarrow \bold{m}+\delta \m$, and use it for the inversion at the next higher frequency.
\ENDFOR
\end{algorithmic}
\end{algorithm}
%%%%%%%%%%%%%%%%%%%%%%%%%%%%%%%%%%%%%%%%%%%%%%%%%%%%%%%%%%%%%%%%%%%%%%%%%%%%%%%%%%%%%%%
%%%%%%%%%%%%%%%%%%%%%%%%%%%%%%%%%%%%%%%%%%%%%%%%%%%%%%%%%%%%%%%%%%%%%%%%%%%%%%%%%%%%%%%

\begin{algorithm}[H]
 \begin{algorithmic}[1]
 \caption{Wavefield-oriented dual AL algorithm for FWI.} \label{alg2}
 \REQUIRE Observed data $\d$, source $\b$, initial model $\m$, fixed data tolerance $\delta_{\omega}$, initial $\mu$
 \FOR{$\omega\in[\omega_{min},\omega_{max}]$}
 \STATE Build $\A_0=\A(\m)$ at the frequency $\omega$ and compute LU factorization of the augmented Helmholtz operator $(\P^T\P+\mu\A_0^T\A_0)=\hat{\bold{L}}\hat{\bold{U}}$ 
 \STATE  Set $\Eps=\bold{0}$
  \FOR{$k=1,2,...,maxit$}
   \STATE      $\u= \hat{\bold{U}}^{-1}\hat{\bold{L}}^{-1}(\P^T\d_{\omega} + \mu \A_0^T(\b_{\omega} - \Eps_{k-1}))$\\
   \STATE  $\lamb = \Eps_{k-1}+\A_0\u - \b_{\omega}$
\STATE  $\delta \m=(\bold{L}(\u)^T\bold{L}(\u))^{-1}\bold{L}(\u)^T \lamb$
\STATE $\Eps_{k}=\Eps_{k-1} + \A(\bold{m}+\delta \m)\bold{u}-\bold{b}_{\omega}$
\ENDFOR
\STATE $\mu= (\|\P\u-\d_{\omega}\|_2+\delta_{\omega})/(2\|\P\u-\d_{\omega}\|_2)\mu$
\STATE Use $\delta\m$ obtained at the current frequency to update the model, $\m \leftarrow \bold{m}+\delta \m$, and use it for the inversion at the next higher frequency.
\ENDFOR
\end{algorithmic}
\end{algorithm}
%%%%%%%%%%%%%%%%%%%%%%%%%%%%%%%%%%%%%%%%%%%%%%%%%%%%%%%%%%%%%%%%%%%%%%%%%%%%%%%%%%%%%%%
\subsection{On penalty parameter selection}
Accurate determination of the penalty parameter $\mu$  is crucial for avoiding the ill-conditioning of the problem and to fit the data at the desired level. At each iteration of the algorithm, the discrepancy principle strategy is employed to automatically determine this parameter, which requires solving the following 1D-root finding problem  \cite{Gholami_2024_GJI}:
\begin{equation}\label{reg_param}
  \phi(\mu):=   \| (  \frac{1}{\mu}  \bold{Q} + \bold{I} )^{-1} \delta \d\|_{2} = \delta_{\omega},
\end{equation}
(line 7 of Algorithm \ref{alg1}) where $\delta_{\omega}$ is an estimate of the norm of noise of data at  frequency $\omega$. 
In this case, the Lagrange multiplier fits the scattered data with the specified noise level. The determination of $\mu$ in the multiplier-oriented algorithm (Algorithm \ref{alg1}) is handled efficiently, even for large-scale problems, as the matrix $\bold{Q}=\S_0\S_0^T$ is sized according to the number of receivers and is computed prior to the iteration. This precomputation allows for careful tuning of $\mu$ at each iteration.
However, in the wavefield-oriented approach, the parameter $\mu$ defines the augmented operator $(\P^T\P+\mu\A_0^T\A_0)$, which is of the size of the model, making it more difficult to adjust $\mu$ at each iteration. For this reason, Algorithm \ref{alg2} employs a fixed value of $\mu$ for each frequency. The value of $\mu$ is determined in line 10 of Algorithm \ref{alg2} based on the method proposed in \cite{Gholami_2022_ABP}.

\subsection{Computational Complexity}
The dual AL method, as presented in Algorithms \ref{alg1} and \ref{alg2}, introduces an efficient alternative strategy that retains the advantages of the original primal/dual AL method while addressing its computational limitations.

\subsubsection{Algorithm \ref{alg1}}
The computational complexity for LU factorization of $\A_0$ is $O(n^3)$, while each forward/backward substitution has a complexity of $O(n^2)$. After computing the LU factorization of $\A_0$, the matrix $\bold{Q}$ requires $2n_r$ forward/backward substitutions, leading to a total cost of $O(2n_r \times n^2)$ floating-point operations. 

Assuming the inversion of $\bold{Q}$ is negligible, the complexity for lines 6, 8, and 9 is dominated by $O(2n_s \times n^2)$. Therefore, the overall complexity of Algorithm \ref{alg1} is driven by the $O(n^3)$ factor, and for $n_{\omega}$ frequencies, the total complexity becomes $O(n_{\omega} \times n^3)$.

\subsubsection{Algorithm \ref{alg2}}
In this case, the LU factorization of $(\P^T\P + \mu\A_0^T\A_0)$ (line 3) has a complexity of $O(n^3)$. At each inner iteration, the wavefield is computed with a complexity of $O(2n_s \times n^2)$ (line 6). Given that these are the dominant operations, for $n_{\omega}$ frequencies, the total complexity is also $O(n_{\omega} \times n^3)$.

Thus, both Algorithms \ref{alg1} and \ref{alg2} have a computational complexity of $O(n_{\omega} \times n^3)$. However, since Algorithm \ref{alg1} allows for more efficient fine-tuning of the penalty parameter $\mu$, it is used for the numerical examples in this paper. Furthermore, Algorithm \ref{alg1} is suitable for the time-domain.

In the standard AL algorithm, the background model is treated as an optimization variable that needs to be updated at each (inner) iteration \cite{Aghamiry_2019_IRWRI}. For $n_{\omega}$ frequencies, this results in a total computational complexity of $O(maxit \times n_{\omega} \times n^3)$. Consequently, the speedup achieved by the proposed dual algorithm over the original primal-dual approach is approximately proportional to $maxit$, representing the number of inner iterations required to invert each frequency.

% \subsection{}
% The primary Lagrange multipliers in \eqref{moalm_lambda} and wavefield in \eqref{moalm_u} require solving $N_r$ and $N_s$ PDEs, respectively, resulting in a total of ($N_r+N_s$) PDE solves during each iteration. The utilization of direct solvers, like LU factorization, can be advantageous in achieving computational efficiency for these multi-right-hand-side problems. For a 2D problem, the total computational complexity required for LU factorization is $O(n^3)$, while for forward/backward substitution in \eqref{moalm_lambda} and \eqref{moalm_u}, it is $O(2n^2.N_r)$ and $O(2n^2.N_s)$. Notably, in \eqref{moalm_lambda}, the matrix $\S$ and subsequently $\S^T$ and $\S\S^T$, as well as $\A^{-1}$ in \eqref{moalm_u}, needed to be computed just once and then utilised for several right-hand side terms.

% Since in dual-AL method, the LU factorization is performed for the background model, the dual-AL algorithm in \eqref{alg1} requires only one LU factorization during the inversion of each data frequency. 
% Having LU factorized matrices, the primary Lagrange multiplier in \eqref{moalm_lambda}, and the extended forward wavefield in \eqref{moalm_u} can be computed by forward/backward substitution, each one scales to $O(n^2)$  meaning a significant reduction in floating-point operations and a fast FWI optimization. 

\section{Interpretation of the Dual method} \label{Interp}
This section gives a brief description of the strategy which the dual algorithm may be interpreted.

Consider the acoustic FWI, which aims to simultaneously find the wavefield $\u$ and the model $\m$ that satisfy both the wave equation $\A(\m)\u  = \b$ and the data equation $\P\u=\d$. Due to the nonlinearity of the problem, iterative linearization is typically applied. Starting from an initial model $\m_0$, the wave equation can be expressed as \cite{Gholami_2024_GJI}:
\begin{equation}
    \A_0\u  = \b - \omega^2 \text{diag}(\delta\m)\u_0 - \omega^2 \text{diag}(\delta\m)\delta\u,
\end{equation}
where $\delta\m=\m-\m_0$, $\delta\u=\u-\u_0$, and $\u_0$ is a known incident wavefield. 

The nonlinear term $-\omega^2 \text{diag}(\delta\m)\delta\u$ on the right-hand side accounts for higher-order scattering and is typically neglected to linearize the equation. In this case, if $\u_0$ satisfies $\A_0\u_0=\b$, the first-order Born approximation is achieved, which forms the basis of conventional FWI algorithms \cite{Pratt_1998_GNF, Virieux_2009_overview}. 
It is also possible to obtain a more accurate incident wavefield that simultaneously satisfies $\A_0\u_0=\b$ and $\P\u_0=\d$ in the least-squares sense, as employed in wavefield reconstruction inversion (WRI) \cite{VanLeeuwen_2013_MLM}:
\begin{equation}
    \begin{pmatrix}
        \beta \A_0\\
        \P
    \end{pmatrix}
    \u_0
     \begin{pmatrix}
        \beta \b\\
        \d
    \end{pmatrix},
\end{equation}
where $\beta>0$ is a balancing parameter. In this case, the approximation goes beyond the first-order Born approximation, resulting in a more accurate linearization \cite{Operto_2023_EFI}. 
In both cases, some higher-order scattering terms are neglected and are subsequently compensated for by updating the model parameters at each iteration.

In order to see how the dual algorithm works, let us substitute $\lamb(\Eps)$ from \eqref{moalm_lambda} into \eqref{moalm_u}, giving
\begin{align}
    \A_0\u(\Eps) = \b + \bold{p} + \bold{q}(\Eps),
\end{align}
where 
\begin{align}
    \bold{p} &=(\bold{S}_0^T\bold{S}_0+\mu \bold{I})^{-1}\bold{S}_0^T\delta \bold{d}_0,\\
    \bold{q}(\Eps)&=-\mu (\bold{S}_0^T\bold{S}_0+\mu\bold{I})^{-1} \Eps.
\end{align}
We observe that the first two terms on the right-hand side are fixed and independent of $\Eps$. It can be shown that $\bold{p}$ corresponds exactly to the source extension used in the WRI method. The last term, $\bold{q}(\Eps)$, is linearly dependent on $\Eps$ and accounts for the higher-order scattering neglected in WRI.

Consequently, unlike traditional methods that compensate for higher-order scattering, the source of the problem's nonlinearity, by iteratively updating the model parameters, the dual algorithm focuses on directly estimating the nonlinear term while keeping the background model fixed. Once this nonlinear term is accurately estimated, the problem becomes fully linearized and can be solved in a single iteration.

\section{Numerical Examples} \label{NumExam}
This section evaluates the computational efficiency and accuracy of the proposed dual-AL method for constant-density (equal to unity) acoustic- and elastic-FWI in isotropic media. For all examples, the model error is computed as 
\begin{equation*}
    \text{ME} (\%)=100 \times \|\m-\m^{*}\|_2 / \|\m^{*}\|_2,  
\end{equation*}
where $\m^*$  denotes the true model.
\subsection{Acoustic inversion}
For acoustic media, the model parameterization is based on the squared slowness ($ \m = \text{V}_\textrm{P}^{-2}$).
The Marmousi II and 2004~BP models are assessed for the purpose validating our inversion algorithm. 
The wave equation is discretized using a nine-point stencil finite-difference technique, which includes the incorporation of an antilumped mass. The stencil coefficients are adjusted based on frequency, as described in \cite{Chen_2013_optimal_FD}. Furthermore, absorbing boundary conditions are implemented along all boundaries.

Throughout the examples, the inversion stage is carried out for specific frequencies within a set of inversion paths, using the established multiscale frequency continuation approach. In this method, the result of each path is used as the starting model for the subsequent path.  
\begin{table*}[]
\caption{Examples and associated acquisition parameters were utilized to gather data regarding the benchmark models.}
\label{Table:models_info}
\resizebox{1\columnwidth}{!}{%
\begin{tabular}{c|ccccccc}
\begin{tabular}[c]{@{}c@{}}Example \\ no.\end{tabular} & \begin{tabular}[c]{@{}c@{}}Inversion \\ type\end{tabular} & \begin{tabular}[c]{@{}c@{}}Model\\  name\end{tabular} & \begin{tabular}[c]{@{}c@{}}Model \\ dimension (km)\end{tabular} & Grid size         & Grid interval (m) & Source no. & Receiver no. \\ \hline
I                                                      & Acoustic                                                  & Marmousi II                                           & 3.5 $\times$ 17.1                                               & 281 $\times$ 1361 & 12.5              & 137        & 137          \\ \hline
II                                                     & Acoustic                                                  & 2004 BP                                               & 12 $\times$ 67.5                                                & 160 $\times$ 900  & 75                & 450        & 67           \\ \hline
II                                                     & Elastic                                                   & SEG/EAGE overthrust                                   & 4.67 $\times$ 20                                                & 187 $\times$ 801  & 25                & 134        & 400          \\ \hline
\end{tabular}%
}
\end{table*}

\subsubsection{Example I: dual-AL versus AL}

\begin{figure}[!h]
\centering
 \includegraphics[scale=1, width=.5\columnwidth]{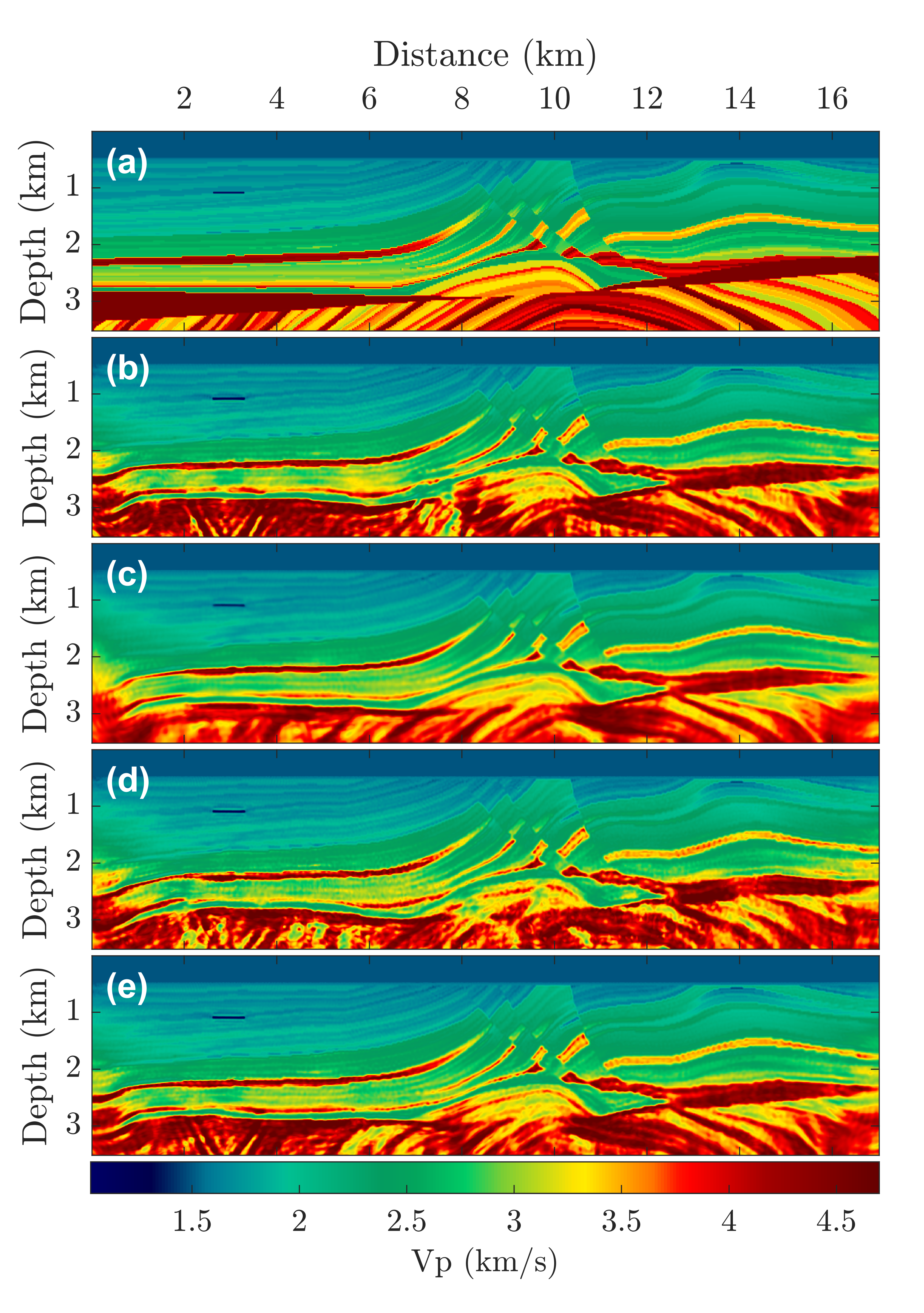}
\caption{The Marmousi test. (a) True model, inversion results achieved by AL (b), dual-AL (c), dual-AL accelerated by l-BFGS with a memory of 10 (d), and dual-AL accelerated by AA with a history of 3 (e).}
\label{Marm_inv_res}
 \end{figure}

The first example focuses on comparing dual-AL and AL algorithms in terms of accuracy and computational efficiency  using Marmousi II velocity model (Fig.~\ref{Marm_inv_res}a). The information regarding the dimension of the model, grid size, grid interval, number of sources and receivers are summarized in Table~\ref{Table:models_info} (first row).
Simulation is performed with a surface acquisition setup using the Ricker wavelet with a dominant frequency of 10~Hz  as the source function.
The inversion process for each method starts with a one-dimensional starting model that exhibits a linear increase from 1.5 km/s to 4.5 km/s. It subsequently takes two paths for inversion, in which frequency range covered by each path is from 3 Hz to 15 Hz, with intervals of 0.5 Hz. 

The AL approach involves a total of 540 iterations. Within these iterations, frequencies of 3 Hz and 3.5 Hz are processed in 20 iterations, while the remaining frequencies undergo inversion in 10 iterations each. This task necessitates the computation of 540 LU factorizations. In contrast, the dual-AL technique requires performing 1 LU factorization per frequency, resulting in a total of 50 LU factorizations. The dual-AL method also includes the same number of iterations to update the dual variable. 

The final inversion results achieved by the AL and dual-AL algorithms are presented in Fig.~\ref{Marm_inv_res}b and Fig.~\ref{Marm_inv_res}c, respectively. The dual-AL approach exhibits remarkable accuracy, despite the substantial redundancy in computational tasks and runtime associated with LU factorizations as summarized in Table.~\ref{Table:AL_vs_DAL}.

\begin{figure}[!h]
\centering
 \includegraphics[width=.5\columnwidth]{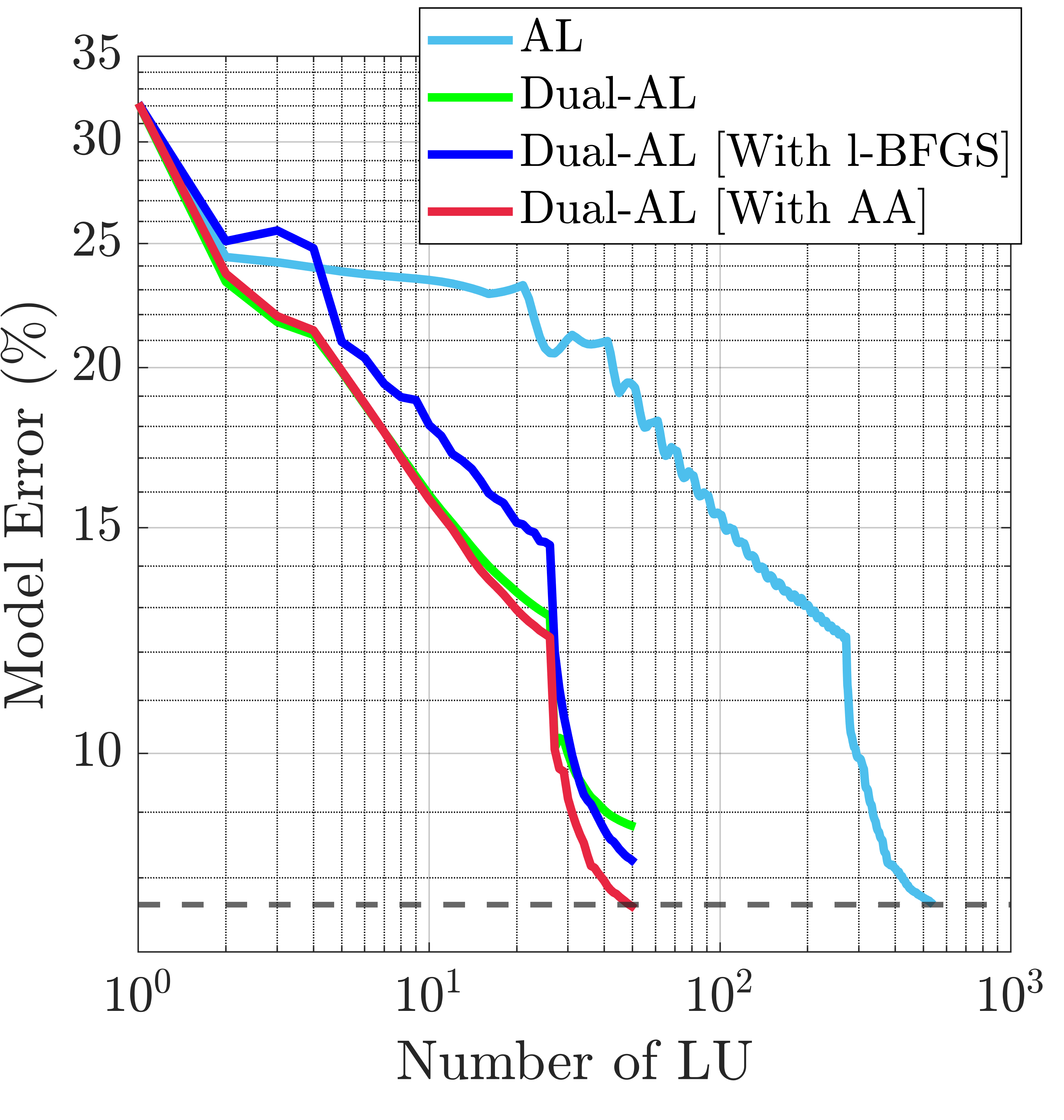}
\caption{The Marmousi test.
The evolution of the computed model error versus number of LU factorization for four different methods: AL, dual-AL, dual-AL by l-BFGS with a memory of 10, and dual-AL by AA with a history of 3.}
\label{marm_error}
 \end{figure}

\begin{figure}[!h]
\centering
 \includegraphics[width=.6\columnwidth]{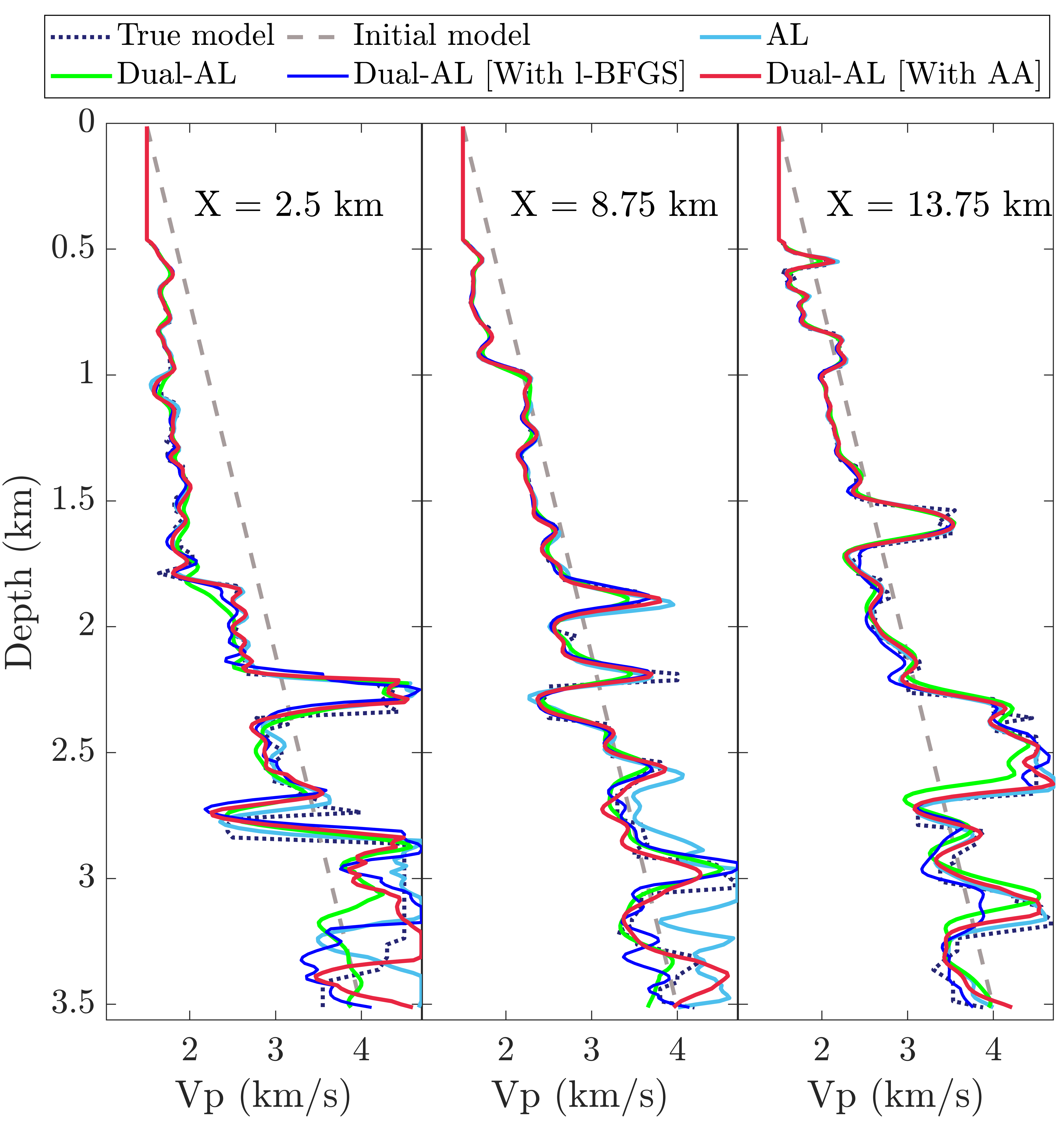}
\caption{Comparison of the vertical velocity profiles extracted from true model, initial model, and the reconstructed models (Figure~\ref{Marm_inv_res}) at different horizontal distances labelled as $X = \text{2.5~km, 8.75~km, and 13.75~km}$.}
\label{marm_logs}
\end{figure}

In order to highlight the importance of the dual update, the l-BFGS (with memory of 10) and AA techniques (with a history of 3) are employed. The reconstructed models of this evaluation are shown in Fig.~\ref{Marm_inv_res}d (for l-BFGS) and Fig.~\ref{Marm_inv_res}e (for AA), each one with an overall higher accuracy compared to the original dual-AL method. However, the accuracy of the dual-AL technique with AA was found to be superior than both original dual-AL and AL methods, as can be verified by the quantitative comparison in terms of computed model errors versus number of LU factorizations (Fig.~\ref{marm_error}). 
Moreover, a well-behaved convergence curve is observed in the case of dual-AL based experiments.
Furthermore, the velocity models that have been reconstructed using the aforementioned methods are compared to the true model at various horizontal distances (Fig.~\ref{marm_logs}). Based on this comparison, the dual-AL approach with AA outperforms the other methods in accurately capturing the deeper parts of the model. Hence, the AA approach is employed in the subsequent examples. 
%
%
%
%
%
%%%%%%%%%%%%%%%%%%%%%%%%%%%%%%%%%%%%%%%%%Table%%%%%%%%%%%%%%%%%%%%%%%%%%%%%%%%%%%%%%%%%%%%%%%%%%%%%%%%

\begin{table}[]
\caption{Quantitative comparison of the dual-AL and AL methods for acoustic FWI of the Marmousi II model.}
\label{Table:AL_vs_DAL}
%\resizebox{1\columnwidth}{!}{%
\centering
\begin{tabular}{l|lll}
Method  & LU no.      & Runtime (h) & ME (\%) \\ \hline
AL      & 540          & 7.75        & 7.62    \\ \hline
dual-AL & 50          & 3.63        & 8.75   
\end{tabular}%
%}
\end{table}

%%%%%%%%%%%%%%%%%%%%%%%%%%%%%%%%%%%%%%%%%%%%%%%%%%%%%%%%%%%%%%%%%%%%%%%%%%%%%%%%%%%%%%%%%%%%%%%%%

\subsubsection{Example II: dual-AL against challenging 2004~BP salt model}
\begin{figure*}[!h]
\centering
 \includegraphics[width=.9\columnwidth]{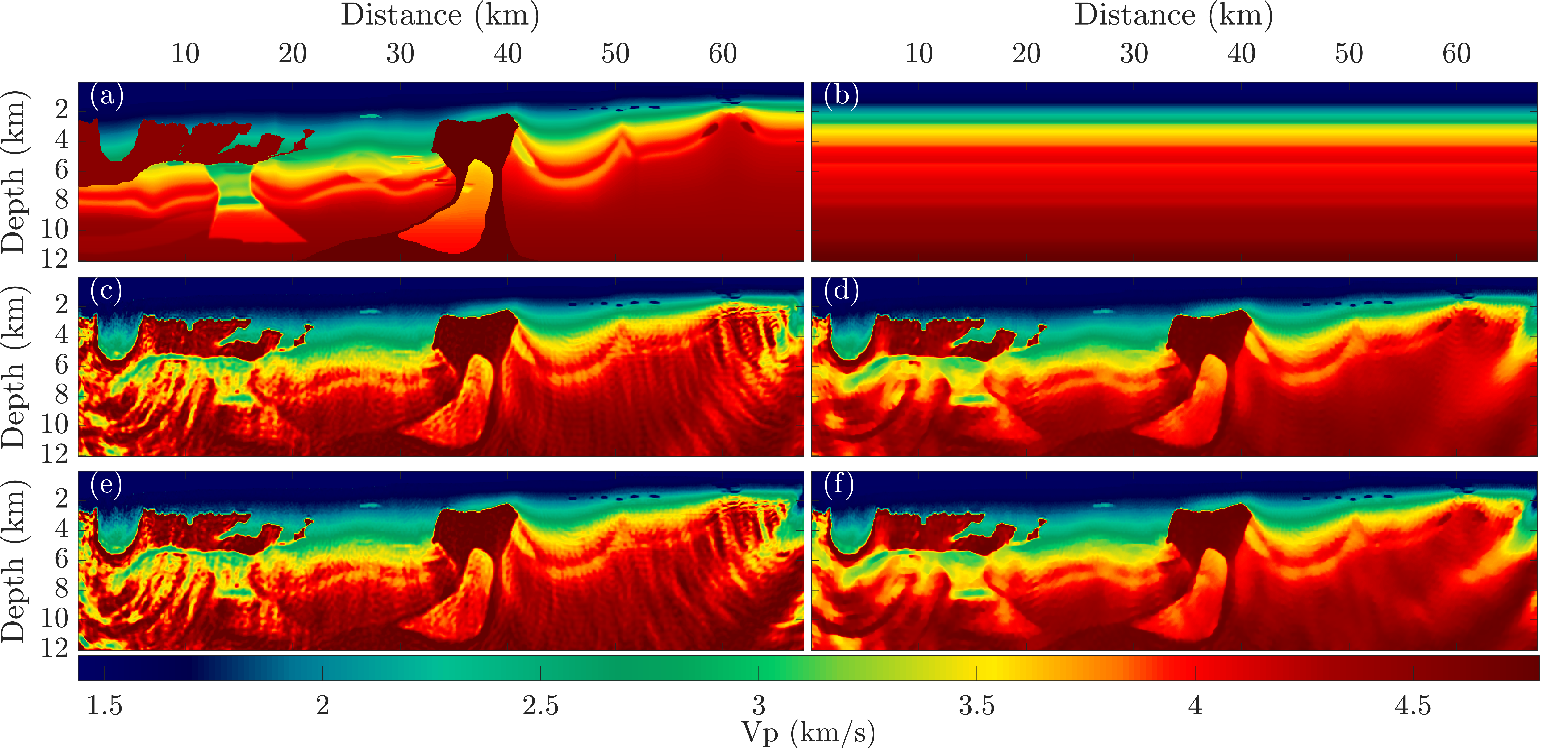}
\caption{2004~BP model example.  (a) True model, and (b) initial model. The inversion results obtained by dual-AL algorithm for the noise-free scenario without regularisation (c) and with TV regularisation (d). (e-f)  same as (c-d) for noise contaminated data.}
\label{BP_inv_res}
 \end{figure*}

The performance of the dual-AL algorithm is evaluated using the challenging 2004~BP salt model (Fig.~\ref{BP_inv_res}a). The information regarding model and acquisition parameters are summarized in Table.~\ref{Table:models_info} (second row). 
An ultra-long-offset stationary-recording acquisition is carried out, in which the pressure sources are 150 m apart at a depth of 25 m and the hydrophones are equally distributed 1 km apart on the seafloor. To enhance computation efficiency, the Green functions spatial reciprocity is utilized to treat sources as receivers and vice versa. The source signature is the Ricker wavelet with a dominate frequency of 3 Hz. 
A one-dimensional initial model was utilized for the inversion (Fig.~\ref{BP_inv_res}b), starting at the data frequency of 1 Hz. According to \cite{sun2020extrapolated}, the above factors presents significant complexities that pose challenges for FWI methods because they often get stuck in a local minimum.

The inversion procedure consisted of four separate paths, each involving the inversion of frequencies within certain ranges. The ranges were 1 Hz to 2.5 Hz, 1 Hz to 3.0 Hz, 1 Hz to 3.5 Hz, and 1 Hz to 4.5 Hz, with a frequency gap of 0.5 Hz inside each. Individual frequency inversions were performed, each with 1 LU factorization and 10 inner iterations to update the dual variable, resulting in a total of 23 LU factorizations. 

\begin{figure}
\centering
 \includegraphics[width=.6\columnwidth]{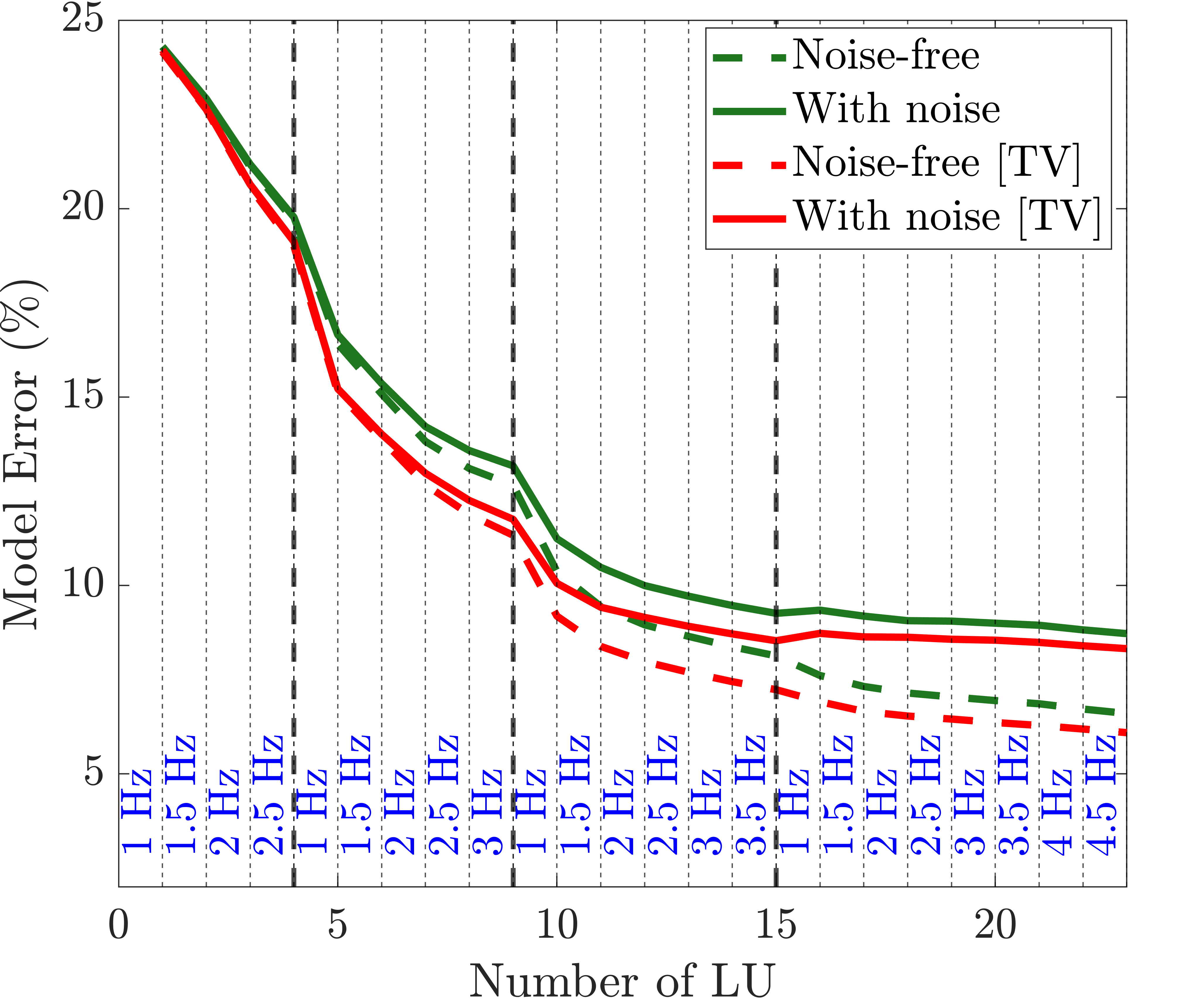}
\caption{2004~BP test. The evolution of the model error versus number of LU factorizations for the estimated models shown in Fig.~\ref{BP_inv_res}.}
\label{BP_MSE}
 \end{figure}

First, the algorithm is evaluated on a noise-free data set. The inversion result shown  in Fig.~\ref{BP_inv_res}c demonstrates the great accuracy of the proposed dual-AL approach, which can be further improved with regularization. Figure.~\ref{BP_inv_res}d illustrates the use of total variation (TV) regularization for suppressing noise-like artifacts. The noise-contaminated data set is next subjected to a similar analysis. To achieve this, Gaussian random noise with a standard deviation equal to 15\% of the mean absolute value of each monochromatic data set is added. Figures~\ref{BP_inv_res}e and \ref{BP_inv_res}f illustrate the inversion results without and with TV regularization, respectively. Compared to noise-free data (Fig.~\ref{BP_inv_res}c), the algorithm maintains resilience in the presence of noise (Fig.~\ref{BP_inv_res}e) thanks to the automatic determination of the regularization parameter in \eqref{reg_param}. A quantitative comparison of the evolution of the model error over the number of LU factorizations is shown in Fig.~\ref{BP_MSE}, confirming the robustness of the proposed method in the presence of noise, as well as the efficacy of TV regularization in improving the results.

\subsection{Elastic inversion}
For elastic media, the model parameterization is based the squared P- and S-wave velocities, $\m  = (\m_\textrm{P},\m_\textrm{S}) \! := (\text{V}_\textrm{P}^2,\text{V}_\textrm{S}^2)$, following \cite{Aghazade_2024_EFWI}.
A 2D section of the 3D SEG/EAGE overthrust is used and the benchmark model to test the proposed inversion algorithm. From this $\text{V}_\textrm{P}$ model, the $\text{V}_\textrm{S}$ model is derived using empirical relation proposed in \cite{Brocher_2005_empirical}, which makes the inversion more difficult because of the existence of various Poisson's ratios within the medium. Forward modeling engine is developed based on the method described in \cite{Chen_2016_elastic_FD}. Furthermore, absorbing boundary conditions are implemented along all boundaries.

Throughout the examples, the inversion stage is carried out for specific frequencies within a set of inversion paths, using the established multiscale frequency continuation approach. In this method, the result of each path is used as the starting model for the subsequent path.  

\begin{figure}[!h]
\centering
 \includegraphics[width=1\columnwidth]{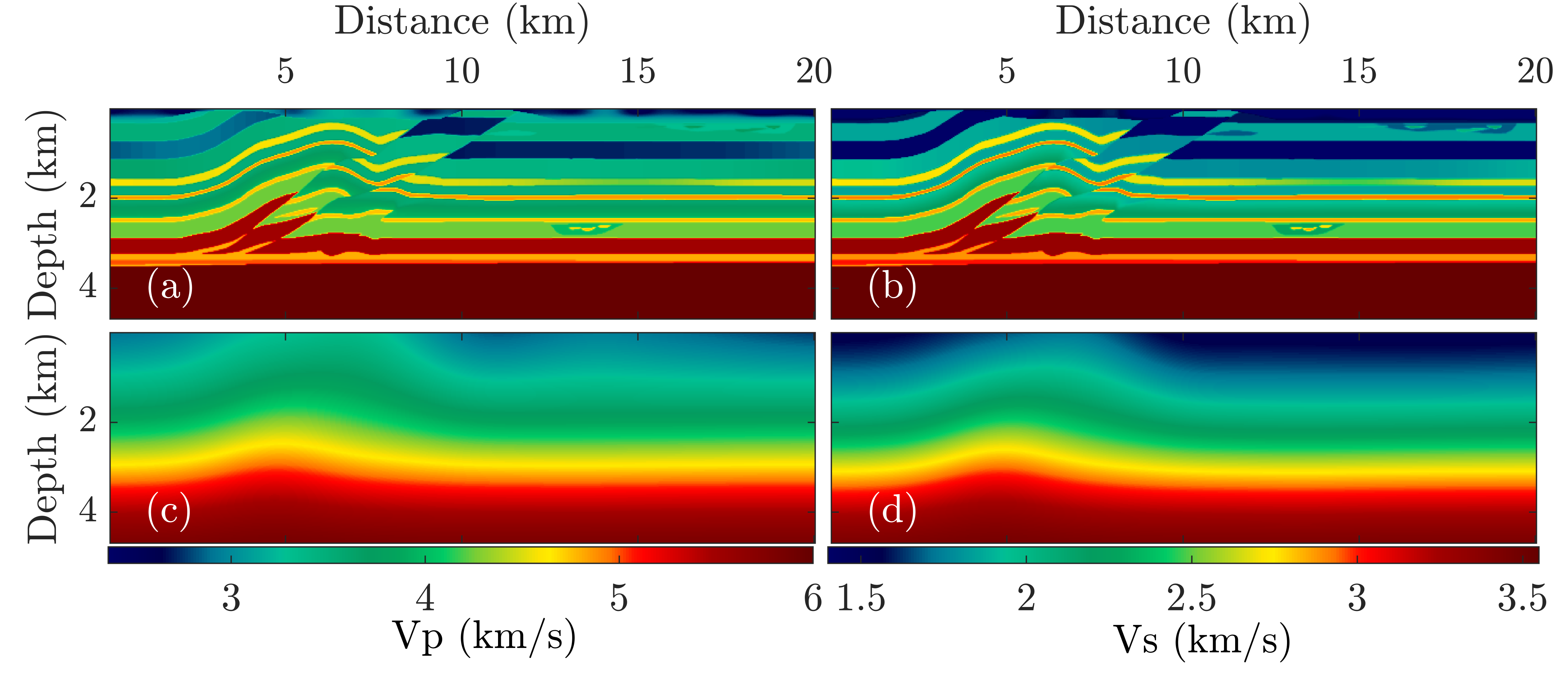}
\caption{Elastic FWI test. (a-b) True $\text{V}_\textrm{P}$ and $\text{V}_\textrm{S}$ models, (c-d) initial $\text{V}_\textrm{P}$ and $\text{V}_\textrm{S}$ models.}
\label{over_true}
 \end{figure}

%\subsubsection{Example III: Elastic-FWI by dual-AL}
%In this section, the performance of the dual-AL method is evaluated for elastic inversion. 
The $\text{V}_\text{P}$ model and empirically derived $\text{V}_\text{S}$ model are shown in Fig.~\ref{over_true}a and Fig.~\ref{over_true}b, respectively with the additional information summarized in Table.~\ref{Table:models_info} (third row).
The surface acquisition setup for the modeling consists of the Ricker wavelet sources (in horizontal and vertical directions) with dominant frequencies of 10 Hz at intervals of 150 m. The receivers are two-component sensors that are positioned at intervals of 50 m. The inversion stage initiated with the initial $\text{V}_\text{P}$ and $\text{V}_\text{S}$ models, respectively shown in Fig.~\ref{over_true}c and Fig.~\ref{over_true}d.

\begin{figure}[!h]
\centering
 \includegraphics[scale=1.4,width=1\columnwidth]{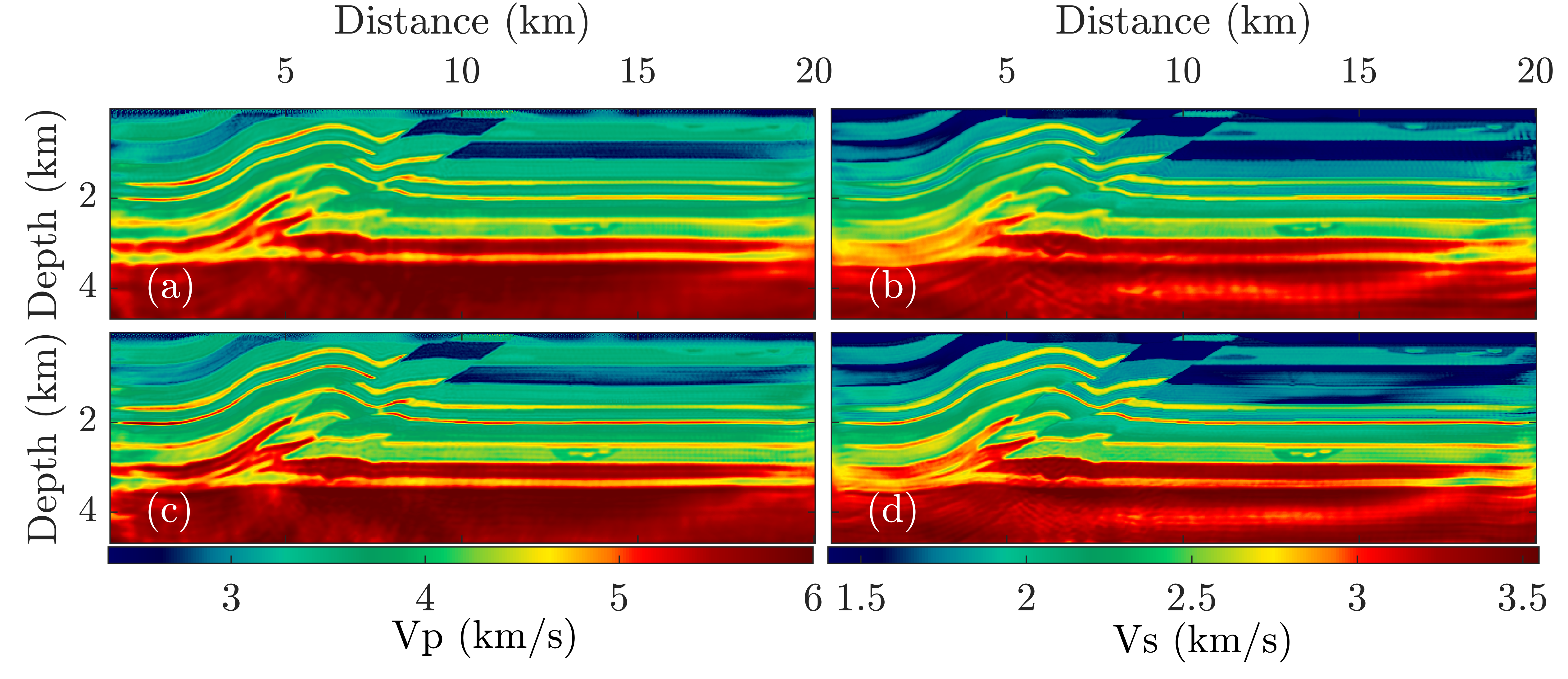}
\caption{Elastic FWI test. (a-b) Reconstructed $\text{V}_\textrm{P}$ and $\text{V}_\textrm{S}$ models obtained by the AL method. (c-d) is the same as (a-b) for the case of dual-AL method.}
\label{over_inv_res}
 \end{figure}

\begin{figure}[!h]
\centering
\includegraphics[width=1\columnwidth]{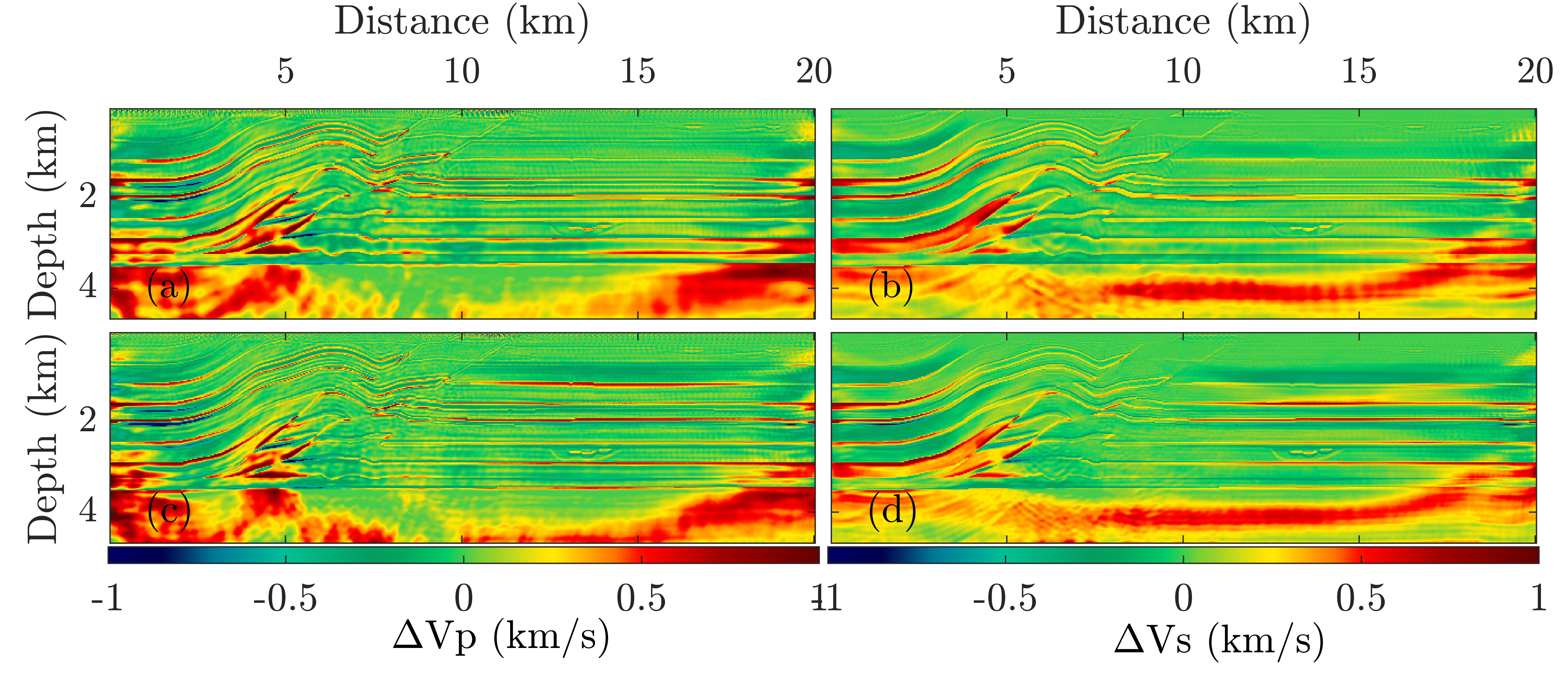}
\caption{Elastic FWI test. The velocity difference between true models shown in Fig.~\ref{over_true}a-b and estimated models shown in Fig.~\ref{over_inv_res}. (a-b) $\text{V}_\textrm{P}$ and $\text{V}_\textrm{S}$ differences for the case of the AL algorithm. (c-d) same as (a-b) for the case of dual-AL algorithm.}
\label{over_inv_diff}
 \end{figure}

\begin{figure}[!ht]
\centering
\includegraphics[width=.7\columnwidth]{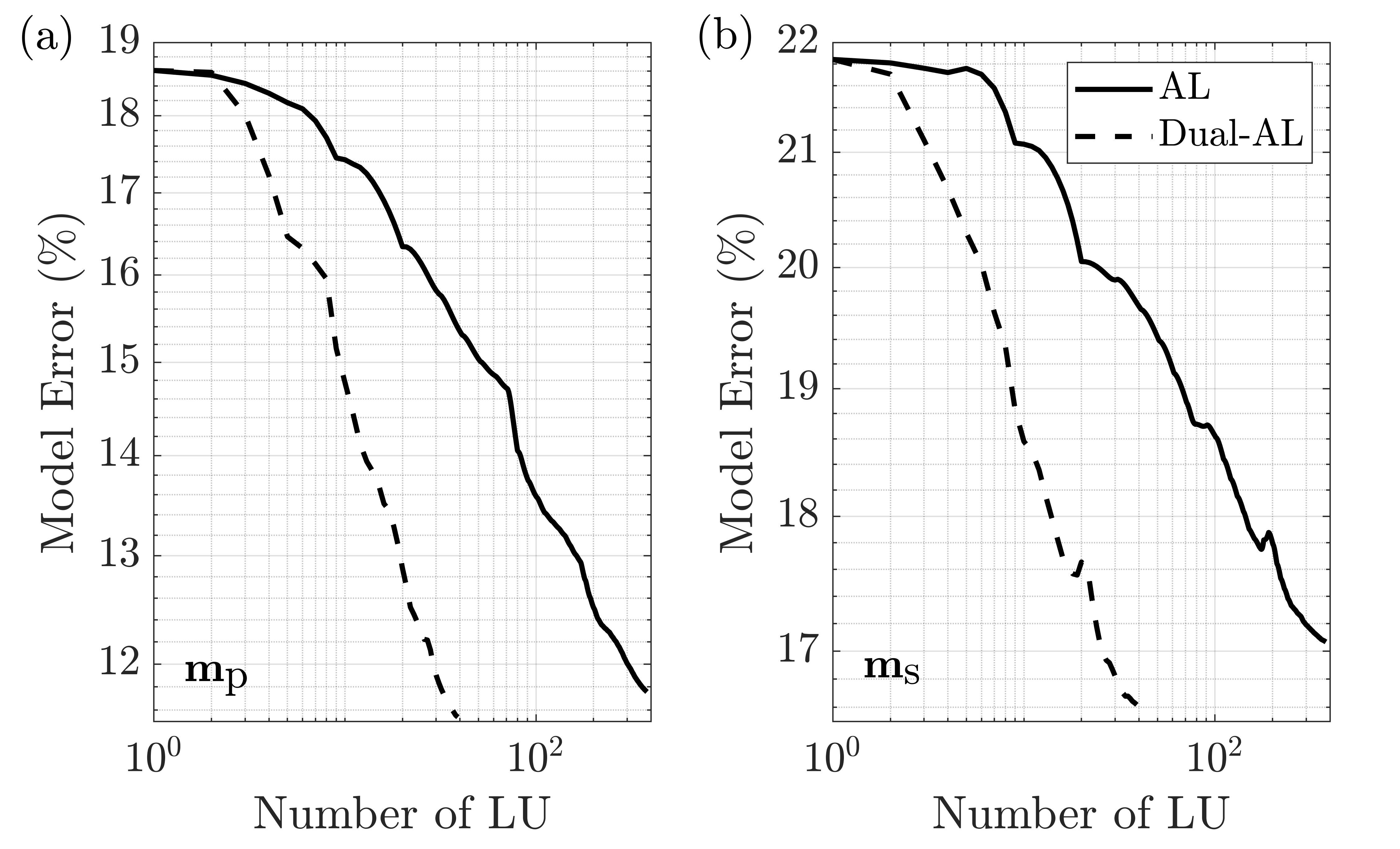}
\caption{Elastic FWI test. Evolution of the computed model error versus the number of LU factorization for (a) squared $\text{V}_\textrm{P}$ model ($\bold{m}_\textrm{p}$) and (b) squared $\text{V}_\textrm{S}$ model ($\bold{m}_\textrm{s}$).}
\label{over_mse}
 \end{figure}

The inversion process consists of three frequency bands: 3 Hz to 6 Hz, 3 Hz to 7.5 Hz, and 3 Hz to 13 Hz, with increments of 0.5 Hz. This results in a total of 38 mono-frequency inversions. Assuming 10 iterations per frequency, a total of 380 iterations were performed.
The performance of the dual-AL algorithm is compared to the original AL method \cite{Aghazade_2024_EFWI}. 
For this example, the AL method requires 380 LU factorizations (one LU factorization per iteration). In contrast, the dual-AL method requires only 38 LU factorizations, with 10 inner iterations per frequency to update the dual variable using AA with a history of 6. The inversion results are shown in Fig.~\ref{over_inv_res}. It can be seen that the reconstructed velocities obtained by the dual-AL method (Fig.~\ref{over_inv_res}c-d) are quite comparable to that of AL (Fig.~\ref{over_inv_res}a-b). This observation is verified by the difference plots (the difference between true and reconstructed velocity models) shown in Fig.~\ref{over_inv_diff}. Overall, the dual-AL method, outperforms the AL approach, which can be verified by the computed model errors shown in Fig.~\ref{over_mse}.

%The inversion technique utilizes a fixed penalty value of $\beta = 10^6$.
%begin{equation}
%\begin{aligned}
%\bold{V}_{S}=&0.7858-1.2344\bold{V}_{P}+0.7949\bold{V}_{P}^2 \\
%&-0.1238\bold{V}_{P}^3+0.0044\bold{V}_{P}^4.
%\end{aligned}
%\end{equation}
%
%
%

 %

 % %%%%%%%%%%%%%%%%%%%%%%%%%%%%%%%%%%%%%%%%%%%%%%%%%%%%%%%%%%%%%%%%%%%%%%%%%%%%%%%%%%%%%%%%%%%%%%%%
%
\section{Discussion} \label{disc}
FWI is a nonlinear optimization problem that involves computationally intensive forward calculations, requiring precise discretization of the associated partial differential equations (PDEs) using methods such as finite differences (FD) to generate synthetic data. Traditionally, the problem's nonlinearity is addressed by expanding around an initial model and truncating higher-order terms to produce a linearized problem, which is then solved iteratively to update the model. 
However, this approach necessitates updating the FD operator and Hessian matrix at each iteration since they depend on the evolving background model. 

In this paper, we proposed an alternative FWI algorithm that maintains a fixed background model and focuses on estimating the missing nonlinearity term associated with it. Our approach is based on solving the dual AL functional and provides an standard formulation for keeping the background model fixed while solving FWI.

This approach of keeping the background model fixed was first introduced in \cite{Abubakar_2009_CSI} within the framework of Contrast Source Inversion (CSI). CSI is based on a quadratic penalty formulation, with optimization performed over both the contrast (wavefield) and contrast source (scattering source). This work served as a source of inspiration for the investigations undertaken in \cite{Alkhalifah_2019_MSF} and \cite{aghamiry2021efficient}. More specifically, \cite{aghamiry2021efficient} applied this approach to the original wavefield-oriented AL method. However, in their formulation, the computed LU decomposition was used as a preconditioner to solve for the wavefield in an inner loop, requiring additional forward/backward substitutions.

Even though the presented dual algorithm leads to a very significant computational savings, its computational efficiency can be further improved by using source encoding  \cite{feng2023multiparameter} or generalized sketching techniques \cite{Aghazade_2021_SWI} to limit the number of right-hand sides.

%Advanced techniques such as the multifrontal massively parallel sparse direct solver (MUMPS, \cite{Amestoy_2006_MUMPS}) are used in FWI \cite{Operto_2007_3Dmodeling,Operto_2015_3DFWI}, along with block low-rank approximation \cite{Amestoy_2016_fast3D_FWI}.

% %%%%%%%%%%%%%%%%%%%%%%%%%%%%%%%%%%%%%%%%%%%%%%%%%%%%%%%%%%%%%%%%%%%%%%%%%%%%%%%%%%%%%%%%%%%%%%%%
%
\section{Conclusion} \label{conc}
We have introduced an efficient algorithm for Full Waveform Inversion (FWI) based on the dual augmented Lagrangian approach. The core innovation of this algorithm lies in its focus on accurately estimating the Lagrange multipliers, which account for the multiple scattering terms inherent to the problem nonlinearity. Once the multipliers are determined, the problem can be effectively linearized and solved with high accuracy for a given background model. This approach allows the finite-difference operator and Hessian matrices to remain fixed throughout the iterations, as they depend solely on the background model. Additionally, the algorithm determines the penalty parameter, its only free parameter, automatically.

Numerical examples using acoustic data demonstrate that the new algorithm can efficiently and automatically reconstruct complex geological models, such as the 2004 BP salt model. The results from elastic data inversion further confirm the algorithm's efficiency and effectiveness in accurately estimating both P-wave and S-wave velocities.

%Constraining the amplitude of the Lagrange multiplier to a specific range enhances the algorithm's resilience and stability. Therefore, future research will be focused on establishing appropriate bounds for Lagrange multipliers along with 3D application. 
%
\section{Acknowledgments}  
This research was financially supported by the SONATA BIS grant
(No. 2022/46/E/ST10/00266) of the National Science Center in
Poland. 
% For peer review papers, you can put extra information on the cover
% page as needed:
% \ifCLASSOPTIONpeerreview
% \begin{center} \bfseries EDICS Category: 3-BBND \end{center}
% \fi
%
% For peerreview papers, this IEEEtran command inserts a page break and
% creates the second title. It will be ignored for other modes.

\bibliographystyle{IEEEtran.bst}

\bibliography{biblio.bib}

%\appendices
%

%

\end{document}